\documentclass{jfm}
\usepackage{amsmath}
\usepackage{amsfonts}
\usepackage{bm}
\usepackage[dvips]{graphicx}

\title{Tumbling of Polymers in a Random Flow with Mean Shear}

\author{M. Chertkov$^1$, I. Kolokolov$^{1,2}$, V. Lebedev$^{1,2}$, and
  K. Turitsyn$^{1,2}$}%

\affiliation{$^1$ Theoretical Division, LANL, Los Alamos, NM 87545, USA
\\[\affilskip]
$^2$Landau Institute for Theoretical Physics, Moscow, Kosygina 2,
  119334, Russia}

\date{\today}

\begin{document}

\maketitle

\begin{abstract}
 A polymer placed in chaotic flow with large mean shear tumbles, making
 a-periodic flips. We describe the statistics of angular orientation, as
 well as of tumbling time (separating two subsequent flips) of
 polymers in this flow. The probability distribution function (PDF) of the
 polymer orientation is peaked around a shear-preferred direction.
 The tails of this angular PDF are algebraic. The PDF of the tumbling time,
 $\tau$, has a maximum at the value estimated as inverse Lyapunov
 exponent of the flow. This PDF shows an exponential tail for large $\tau$ and a
 small-$\tau$ tail determined by the simultaneous statistics of velocity PDF.
\end{abstract}


\underline{\bf Introduction.} With the development of novel optical methods a number of high
quality experimental observations resolving individual polymer (e.g. DNA molecule) dynamics for
elongational and shear regular flows have been reported by \cite{97PSC}, \cite{98SC} and by
\cite{99SBC} and \cite{01HSBSC}, respectively. Experimental explorations by \cite{99SBC} and the
subsequent theoretical/numerical study by \cite{00HSL} of the shear flow setting had focused on the
analysis of the power spectral density and the simultaneous statistics of polymer extension, for
the case in which fluctuations are driven by thermal noise. Additionally, one noticeable
phenomenon, called tumbling, was reported by \cite{99SBC} for the regime of strong shear flow:  The
molecule which spends most of the time being oriented along the direction dictated by the shear
sometimes and suddenly (a-periodically) flips. In another experimental breakthrough, a chaotic flow
state called by the authors ``elastic turbulence'' was observed for dilute polymer solutions by
Groisman \& Steinberg (2000,2001,2004). This flow consists of regular (shear-like) and chaotic
components, the latter being weaker. Resolving an individual polymer in this steady chaotic flow
was more challenging than in the regular flow experiment but still it appeared to be an accessible
task for \cite{04GCS}. The coil-stretch transition, predicted by Lumley (1969,1973) (see also
Balkovsky et al. (2000,2001) and \cite{00Che}), was observed, for the first time, in direct
single-polymer measurements by \cite{04GCS}. The statistics of the polymer extension and of the
tumbling time were also tested in the elastic turbulence experiments, however the phenomenon has
not yet been fully explored.

In this letter, we discuss the statistics of polymers placed in a chaotic flow with a
relatively large mean shear, that is the flow of the type correspondent to the elastic
turbulence experiments by Groisman \& Steinberg (2000,2001,2004). We assume that the
effect of velocity fluctuations is stronger than that related to thermal noise and that
polymers are essentially elongated due to the fluctuations so that the polymer
orientation is well defined. The main body of the orientational fluctuations occur in a
neighborhood of a special direction preferred by the shear. Sometimes these typical
fluctuations around the preferred direction are interrupted by flips, in which the
polymer orientation is reversed. The task of this study is to describe the statistics of
the angular orientation and tumbling time.

The structure of this letter is as follows. We begin by introducing the basic dumb-bell-like
equation governing the dynamics of the polymer end-to-end vector in a non-homogeneous flow. If the
effect of thermal fluctuations is negligible, the angular part of the polymer dynamics decouples
from its extensional counterpart and can be examined separately. It is convenient to count the
angular degrees of freedom $\phi$ and $\theta$ (for in-plane and off-plane orientations,
respectively) from the direction prescribed by the shear. We show that the angular PDF is peaked at
some small angle $\phi$, estimated by the average value $\phi_t=\langle\phi\rangle$, $\phi_t\ll 1$.
The widths (in both angles) of the main part of the PDF are of the order of $\phi_t$. Then we
demonstrate that tails of the joint PDF are algebraic at $\phi_t\ll\phi,\theta\ll 1$. We find that
this algebraic tail of the individual PDF of $\phi$ is related to purely deterministic (i.e. shear
driven) dynamics: ${\cal P}(\phi) \propto\sin^{-2}\phi$. The tail of the $\theta$ PDF has two
competing contributions, one related to deterministic dynamics, $\propto \theta^{-2}$ for
$\phi\ll|\theta|\ll1$, and the other one related to stochastic dynamics, $\propto \theta^{-a}$,
where $a$ is a number, dependent on details of the velocity fluctuation statistics. Then we examine
the statistics of the tumbling time, $\tau$, that is defined as the time between two subsequent
flips of the polymer. The PDF of $\tau$ is peaked at a time estimated by the inverse Lyapunov
exponent of the flow, $\tau_t=\bar\lambda^{-1}$. The long time, $\tau\gg\tau_t$, tail of the PDF is
exponential, $\ln[P(\tau)]\sim -\tau/\tau_t$. The statistics of small tumbling times is related to
the simultaneous PDF of the velocity gradients. To derive these results we explore the close
relation between the stochastic dynamics of $\theta$ and $\phi$ and Lagrangian dynamics of the
flow.separation in the flow. We conclude by discussing the applicability conditions for our
approach and the validity of the assumptions made in this letter.

\underline{\bf Model.} We consider a single polymer molecule which is advected by a
chaotic/turbulent flow. We assume that the velocity correlation length is much larger
than the size of the polymer. (Note, that this condition is always satisfied in elastic
turbulence simply because the velocity correlation length coincides with the overall size
of the flow/apparatus that is the biggest scale in the problem.) Then the polymer can be
viewed as a material point moving along a Lagrangian trajectory. In addition, the polymer
is stretched due to the flow inhomogeneity. The polymer stretching can be characterized
by its end-to-end separation vector, ${\bm R}$. The stochastic dynamics of the vector
${\bm R}$ can be examined in the framework of the following dumb-bell-like equation (see
e.g. \cite{77Hin} and \cite{87BCAH})
 \begin{equation}
 \partial_t R_i=R_j\nabla_j v_i-\gamma R_i +\zeta_i\,,
 \label{basic} \end{equation}
where $\gamma$ is the polymer relaxation rate (dependent on $R$, the absolute value of
the vector $\bm R$), the velocity gradient $\nabla_j v_i$ is taken at the molecule
position, and $\zeta_i$ is the Langevin force. The velocity difference between the end
points of the polymer is approximated in Eq. (\ref{basic}) by the first term of its
Taylor expansion in the end-to-end vector. This is justified by the smallness of the
polymer extension $R$ in comparison with the velocity correlation length.

We focus on the situation for which the effect of velocity fluctuations is stronger than
the effect of thermal noise. Then the Langevin force in Eq. (\ref{basic}) can be
neglected. In this case the polymer angular (orientational) dynamics described by the
unit vector $\bm n=\bm R/R$ decouples from the dynamics of the end-to-end polymer length
$R$ and one derives from Eq. (\ref{basic}) a closed equation for $\bm n$:
 \begin{equation}
 \partial_t n_i=n_j(\delta_{il}-n_in_l)\nabla_j v_l  \,.
 \label{direct} \end{equation}
Note that Eq. (\ref{direct}) coincides with the dynamics of a rod-like micro-object immersed in the
same flow.

 \begin{figure}
 \includegraphics[width=0.6\textwidth]{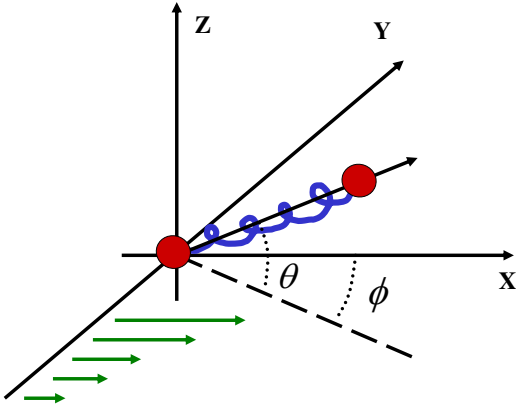}
 \caption{Schematic figure explaining polymer orientation geometry.}
 \label{fig:orient}
 \end{figure}

Let us choose the coordinate frame associated with the mean shear velocity, as shown in
Fig. \ref{fig:orient}. In this reference frame the mean flow is characterized by the
velocity $(sy,0,0)$ where $s$ is assumed to be positive. The polymer orientation is
conveniently parameterized in terms of the angles $\phi$ and $\theta$:
$n_x=\cos\theta\cos\phi$, $n_y=\cos\theta\sin\phi$, $n_z=\sin\theta$. Then, Eq.
(\ref{direct}) becomes
 \begin{eqnarray} &&
 \partial_t{\phi} = - s\sin^2\phi + \xi_\phi \,,
 \label{phieq} \\ &&
 \partial_t{\theta} =- s
 \sin\phi\cos\phi\sin\theta\cos\theta
 + \xi_\theta \,,
 \label{thetaeq} \end{eqnarray}
where $\xi_\phi$ and $\xi_\theta$ are random variables related to the fluctuation
component of the velocity gradient.

\underline{\bf Angular statistics.} The statistics of the velocity
fluctuations is assumed to be homogeneous in time. In a
statistically stationary velocity field, the angular statistics is
stationary as well, being characterized by the joint PDF, ${\cal
P}(\phi,\theta)$, which is a periodic function of the angles with
the period $\pi$ for both $\phi$ and $\theta$. Thus, it is
sufficient to consider ${\cal P} (\phi,\theta)$ within the
following bounded domain (torus), $-\pi/2<\phi,\theta<\pi/2$. We
normalize the PDF using $\int d\phi\,d\theta\,{\cal
P}(\phi,\theta)=1$, where the integral is taken over the domain.
Note, that according to the structure of Eqs.
(\ref{phieq},\ref{thetaeq}), $\cal P(\phi,\theta)$ is symmetric
with respect to $\theta$ but it is not symmetric with respect to
$\phi$. Therefore, the average value of $\phi$,
$\phi_t=\langle\phi\rangle$, is non-zero. In our setting, $\phi_t$
is positive. The value of $\phi_t$ can be estimated by balancing
the deterministic and stochastic terms on the right hand side of
Eq. (\ref{phieq}). The weakness of the random term in comparison
with $s$ implies $\phi_t\ll 1$. The same quantity $\phi_t$
estimates typical fluctuations of $\phi$ about its mean value.
Once one assumes that the random terms in Eqs.
(\ref{phieq},\ref{thetaeq}) are comparable, it immediately follows
that typical value of $\theta$ fluctuations is estimated by
$\phi_t$ as well.

Note that the equation, $\partial_t r_i=r_j\nabla_j v_i$, describing dynamics of separation between
two neighboring fluid particles (moving along nearby Langangian trajectories), leads to the same
dynamics for the unit vector $\bm r/r$ as determined by Eq. (\ref{direct}) and, consequently, to
the same angular dynamics as described by Eqs. (\ref{phieq},\ref{thetaeq}). For $r$ (the absolute
value of ${\bm r}$) one derives: $\partial_t\ln r=s\cos^2\theta \cos\phi\sin\phi +\xi_\parallel$,
where $\xi_\parallel$ represents the direct (as opposed to indirect through fluctuations in $\phi$)
effect of velocity fluctuations. It follows from Eq. (\ref{phieq}) that for typical fluctuations
(when $\phi,\theta\ll1$) $\xi_\phi$ competes with $s\phi^2$. Assuming that $\xi_\parallel\sim
\xi_\phi$ one finds that $\xi_\parallel$ is negligible in comparison with $s\phi$, because the
relevant values of $\phi$ are small, $\phi\ll 1$. Therefore, for small $\phi,\theta$, one arrives
at $\partial_t\ln r= s\phi$. This equation establishes the relation between the angular dynamics of
the polymer and the dynamics of Lagrangian separation. For the Lyapunov exponent, $\bar\lambda$,
defined as the mean logarithmic rate of divergence of Lagrangian trajectories, one finds
$\bar\lambda= \langle\partial_t \ln r \rangle= s\phi_t$.

It is natural to expect that the Lagrangian velocity correlation time is
$\tau_t=\bar\lambda^{-1}=(s\phi_t)^{-1}$, that is also characteristic time of the
$\xi_\phi$ and $\xi_\theta$ fluctuations. Then, comparing the left hand sides of Eqs.
(\ref{phieq},\ref{thetaeq}) with the first terms on their right hand sides (for
$\phi,\theta\ll 1$), one concludes that the angular correlation time can be estimated by
the same quantity $\tau_t$. Next, equating the terms on the right hand sides of Eqs.
(\ref{phieq},\ref{thetaeq}), one derives $\xi_\phi\sim\xi_\theta \sim s\phi_t^2\ll s$.
The last inequality reflects the assumed weakness of the velocity gradient fluctuations
compared to the shear rate, $s$.

\underline{\bf Tails of the angular PDFs.} Let us consider the domain
$|\phi|,|\theta|\gg\phi_t$, where the random terms in Eqs. (\ref{phieq},\ref{thetaeq}),
$\xi_\phi$ and $\xi_\theta$, are negligible. The angular dynamics is purely deterministic
in this domain leading to the following dependence of the angles on time $t$
 \begin{equation}
 \cot\phi=s(t-t_0)\,, \qquad
 \tan\theta= c \cdot\sin\phi \,,
 \label{deter2} \end{equation}
where $t_0$ and $c$ are some constants. According to Eq. (\ref{deter2}), the vector $\bm
n$ reverses its direction as $t$ increases. Therefore, Eq. (\ref{deter2}) describes a
single flip of the polymer. Due to the assumed homogeneity in time of the velocity
statistics, $t_0$ is homogeneously distributed. Recalculating the measure $dt_0$ into the
PDF of the angles in accordance with Eq. (\ref{deter2}), one derives
 \begin{equation}
 {{\cal P}}(\phi,\theta)
 =\frac{U(\tan\theta/\sin\phi)}{\sin^3\phi\,\cos^2\theta} \,.
 \label{deter} \end{equation}
The function $U$ reflects possible variations in $c$ (its statistics), which should be
determined from the initial conditions for deterministic evolution. These conditions have
to be found from matching Eq. (\ref{deter2}) to those defined for the stochastic domain
$|\phi|,|\theta|<\phi_t$. One concludes that the function $U$ is sensitive to the angular
dynamics in the stochastic domain and, respectively, to details of velocity fluctuations,
i.e. the function is nonuniversal. Note, that Eq. (\ref{deter}) is identical to the one
found in \cite{72HL} in the context of a solid rod tumbling in shear flow caused by
thermal (Langevin) fluctuations.

Eq. (\ref{deter2}) shows that in the deterministic regime the angle $\phi$ decreases uniformly with
time (that is except for the jump from $-\pi/2$ to $+\pi/2$ at $t=t_0$). Therefore, the stationary
PDF for the angular degrees of freedom, ${\cal P}(\phi,\theta)$, corresponds to a non-zero
probability flux from positive to negative $\phi$ related to a preferred (clock-wise) direction of
the polymer's rotations in the $X-Y$ plane. Formally, the probability flux goes out through
$\phi=-\pi/2$ and the same flux comes back (enters) through $\phi=\pi/2$ ($\pi/2$ and $-\pi/2$ are
identical by our construction) thus keeping the total probability equal to unity.

The PDF of $\phi$, $P_\phi$, can be obtained from the joint PDF: $P_\phi=\int
d\theta\,{\cal P}(\phi,\theta)$. Integrating the right-hand side of Eq. (\ref{deter})
over $\theta$ one obtains the following expression for the tail, valid for
$|\phi|\gg\phi_t$:
 \begin{equation}
 P_\phi\equiv\int d\theta\,{{\cal P}}(\phi,\theta)
 =C \phi_t \sin^{-2}\phi \,,
 \label{deter1} \end{equation}
where the constant $C$ is of order unity. Let us reiterate that, thinking dynamically,
Eq. (\ref{deter1}) originates from the deterministic flips bringing $\phi$ from its most
probable domain $\sim \phi_t$ to the observation angle. Eq. (\ref{deter1}) describes the
aforementioned probability flux: as determined by Eq. (\ref{phieq}), $P_\phi
\partial_t\phi$ is constant in the deterministic region.

Consider the PDF of $\theta$, $P_\theta=\int d\phi\,{\cal P} (\phi,\theta)$. The naive
result for the PDF tail following from Eq. (\ref{deter}) is
$P_\theta\propto|\theta|^{-2}$, provided $1\gg\theta\gg\phi$. However, one should be
careful, since the expression (\ref{deter}) does not cover a special angular domain,
characterized by $|\phi|<\phi_t$ and $|\theta|\gg \phi_t$, which should be analyzed
separately. In this domain, one can neglect $\xi_\theta$ in Eq. (\ref{thetaeq}). Assuming
also $|\theta|\ll1$, one arrives at $\partial_t\ln(\theta)=-s\phi$. In this case, $s\phi$
can be treated as a random variable independent of $\theta$ and the above equation leads
to an algebraic tail, ${\cal P}\propto |\theta|^{-a}$, where the exponent $a$ is a
positive number of order unity. The value of $a$ is sensitive to the statistics of the
$\phi$ fluctuations. (Therefore, $a$ is not universal.) Let us explain the origin of the
algebraic dependence. The algebraic contribution to the angular PDF is related to the
long (compared to the correlation time $\tau_t$) period when $\phi$ fluctuates around
some negative value, $\sim-\phi_t$. (These fluctuations can be interrupted by flips.)
Then $\theta$ at the end of the $T$-long period is estimated according to
$\ln(\theta/\phi_t)\sim T s \phi_t$. The probability $W$ to observe such a long a-typical
period is estimated by $\ln W \sim -T/\tau_t$. These estimates, recalculated in the PDF
of $\theta$, $P_\theta=dW/d\theta$, give the aforementioned algebraic tail. Note that
this algebraic tail which originates from the long-time dynamics is analogous to the
algebraic tail of the polymer extension PDF discussed by Balkovsky et al. (2000,2001) and
\cite{00Che}.

Therefore, one finds that there exist two different contributions to the PDF tail: one related to
the deterministic motion, described by Eq. (\ref{deter2}), while the other is associated with the
stochastic evolution in the domain, $|\phi|<\phi_t,|\theta|\gg \phi_t$. For
$1\gg|\theta|\gg\phi_t$, both contributions are algebraic, $\propto|\theta|^{-2}$ and
$\propto|\theta|^{-a}$, respectively. The deterministic contribution, $\propto|\theta|^{-2}$,
dominates if $a>2$, while the stochastic contribution, $\propto|\theta|^{-a}$, dominates otherwise.

\underline{\bf Tumbling time statistics.} As seen from the expression (\ref{deter2}), the
deterministic process, which actually defines the polymer turn (because $\phi$ changes
essentially only during deterministic part of the dynamics), is faster than the
stochastic wandering taking place at small angles, $|\phi|,|\theta|<\phi_t$. Therefore it
is convenient to define the tumbling time, $\tau$, as the time separating two subsequent
crossings in $\phi$ of the special angle $\pm\pi/2$, in the middle of the deterministic
domain.  Since the major contribution to $\tau$ originates from the stochastic wandering
in the $\phi_t$-narrow vicinity of $\phi=0$, the position of the $\tau$-PDF maximum and
its width are both estimated by the correlation time $\tau_t= (s\phi_t)^{-1}$, because
this is the only relevant characteristic time of the stochastic angular evolution.

Being interested in the PDF tail, for $\tau\gg\tau_t$, one observes that if a flip does
not occur for a long time, then this delay can be interpreted in terms of the large
number, $\tau/\tau_t$, of independent unsuccessful attempts to pass (clock-wise in
$\phi$) the stochastic domain $|\phi|<\phi_t$. The probability of the delayed flip is
given by the product of the probabilities of these $\tau/\tau_t$ events, resulting in the
exponential tail of the PDF of $\tau$ for $\tau\gg\tau_t$, $\ln P_\tau\sim -\tau/\tau_t$.

The left, $\tau\ll\tau_t$, tail of the tumbling time PDF is non-universal because it is sensitive
to details of the velocity field statistics. Indeed, it is determined by the special configurations
of the velocity field that force $\phi$ to drift through the stochastic region a-typically fast.
(Those configurations are vortices with clock-wise rotation of the fluid in the $X-Y$ plane leading
to negative values of $\xi_\phi$ that are larger than $s\phi_t^2$ in absolute value.) Analyzing the
anomalously fast revolutions of the polymer, one finds that $\xi_\phi$ from the right hand side of
Eq. (\ref{phieq}) may be considered as time-independent. (Here again, the natural assumption is
that the correlation time of the velocity field fluctuations is of the same order as the inverse
Laypunov exponent in the flow, i.e. $\sim\tau_t$.) Then the direct solution of Eq. (\ref{phieq})
gives $\tau=\pi/\sqrt{|\xi_\phi|s}$, where we assumed that the major contribution in $\tau$ comes
from the domain of small $\phi$, $\phi\ll 1$. This estimate holds if $s\gg|\xi_\phi|\gg s\phi_t^2$.
For $1/s\ll\tau\ll\tau_t$ one arrives at the following expression for the PDF of $\tau$:
 \begin{equation}
 P_\tau=\frac{2\pi^2}{\tau^3s}
 P_\xi\left(-\frac{\pi^2}{\tau^2s}\right) \,,
 \label{small_tau} \end{equation}
where $P_\xi$ is the single-time PDF of $\xi_\phi$.

\underline{\bf Conclusions.} Let us discuss applicability conditions of our results. One of our
assumptions was that the mean flow can be approximated by a perfect shear flow, whereas in reality
flow parameters vary along the Lagrangian trajectory demonstrating essential spatial
inhomogeneities. Even though these variations were not included in our derivations, our results
remain valid if the variations along the Lagrangian trajectory occur on time scales larger than
$\tau_t$ and also if the local flow does not deviate strongly from the shear configuration. Then,
the PDFs discussed in this paper adjust adiabatically to the current values of the parameters and
become, consequently, slow functions of spatial position. If the regular part of the flow is
elongational, polymer flips become forbidden in the ideally deterministic regime while fluctuations
will still generate some tumbling. Analysis of the tumbling in elongational, and, generically in
any other random flow, will be discussed elsewhere.

We have focused on discussing the dynamics and the statistics of the polymer's angular
degrees of freedom.  However, our theoretical scheme can be naturally extended to also
include polymer extension. The stochastic dynamics and the statistical properties of the
polymer extension can be examined in a way very similar to the one developed in this
letter and this will be the subject of separate publication.

Our theory is based on the simple dumb-bell-like equation (\ref{basic}).  This equation
is obviously approximate, taking into account only one variable (end-to-end vector ${\bm
R}$) of generally more complex dynamics. Therefore, it should be important to assess the
effects of more realistic modeling. (This treatement should also account for internal
conformational degrees of freedom.)

Note that polymer tumbling was first observed in the steady shear flow experiments of
\cite{99SBC} and \cite{01HSBSC} in which orientational fluctuations were driven by
thermal noise, while our analysis has focused primarily on the case of tumbling driven by
velocity fluctuations. Therefore, even though all of our results are directly applicable
to the elastic turbulence setting of Groisman \& Steinberg (2000,2001,2004), the
examination of the statistics of the Langevin driven tumbling and angular distribution
is, actually, a separate task. In particular, special attention should be given to the
case of extremely elongated polymers in which the non-linearity of the polymer elasticity
is essential. These Langevin-related problems will be examined elsewhere. Here, let us
only note the universal nature of the exponential, large $\tau$, tail of the tumbling
time PDF. Indeed, it is straightforward to check that the arguments presented above are
generic, thus guaranteeing that, very much like in the velocity fluctuation driven case,
the tail is also exponential in the Langevin-driven case.

We thank V. Steinberg for many stimulating discussions and G. D. Doolen for useful
comments. We acknowledge support of RSSF through personal grant (IK), of RFBR, grant
04-02-16520a (IK,VL and KT), and of Dynasty Foundation (KT).

\end{document}